\date{}
\documentclass[aps,pra,showpacs,twocolumn]{revtex4-1}
\usepackage{graphicx,psfrag,amsmath,amssymb,amsfonts,latexsym,color,dcolumn}
\begin{document}
\title{Vacuum fluctuations and generalized boundary conditions}
\author{C\'esar D. Fosco$^{1,2}$}
\author{Fernando C. Lombardo$^3$}
\author{Francisco D. Mazzitelli$^{1}$}
\affiliation{$^1$ Centro At\'omico Bariloche,
Comisi\'on Nacional de Energ\'\i a At\'omica,
R8402AGP Bariloche, Argentina}
\affiliation{$^2$ Instituto Balseiro,
Universidad Nacional de Cuyo,
R8402AGP Bariloche, Argentina}
\affiliation{$^3$ Departamento de F\'\i sica {\it Juan Jos\'e
 Giambiagi}, FCEyN UBA, Facultad de Ciencias Exactas y Naturales,
 Ciudad Universitaria, Pabell\' on I, 1428 Buenos Aires, Argentina - IFIBA}
\date{today}
\begin{abstract} 
We present a study of the static and dynamical Casimir effects for a
quantum field theory satisfying generalized Robin boundary condition, of a
kind that arises naturally within the context of quantum circuits.  Since
those conditions may also be relevant to measurements of the dynamical
Casimir effect, we evaluate their role in the concrete example of a real
scalar field in $1+1$ dimensions, a system which has a well-known
mechanical analogue involving a loaded string.  
\end{abstract}
\pacs{12.20.Ds, 03.70.+k, 11.10.-z}
\maketitle
\section{Introduction}\label{sec:intro}
Over the course of the last 15 years, there has been a renewed interest in
the Casimir effect. This was partly due to a second generation of
experiments, started in 1997, which triggered a sustained flow of both
theoretical and experimental works.  As an outcome of those works, our
knowledge about the dependence of Casimir forces on the geometry and
material properties of the objects involved has remarkably 
improved~\cite{libros}.  

The dynamical counterpart of the Casimir effect, also known as `motion
induced radiation', has also been the subject of intense research,
manifested in the consequent profusion of works~\cite{reviews dce}.
Although the direct measurement of radiation generated by moving mirrors is
a daunting experimental challenge, photon creation induced by time
dependent external boundary conditions has, indeed, been observed
experimentally, albeit in a different context, namely, superconducting
circuits~\cite{wilson1}.  There are also ongoing experiments aimed at
measuring the photon creation induced by the time dependent conductivity of
a semiconductor slab enclosed by an electromagnetic cavity~\cite{padova},
and proposals based on the use of high frequency resonators to produce the 
photons,  and ultracold atoms to detect the created photons via superradiance
\cite{Kim06}. 

The superconducting circuit experiment mentioned above~\cite{wilson1},
consists of a coplanar waveguide terminated by a SQUID, upon which a time
dependent magnetic flux is applied. This system can be described by a
quantum scalar field $\phi(x,t)$ (the magnetic flux at the different
positions of the transmission line) satisfying on the SQUID (located at $x=a$) the boundary
condition~\cite{Johansson}
\begin{equation}\label{eq:gbc}
\frac{1}{L}\frac{\partial\phi}{\partial x}(a,t)+C\frac{\partial^2
\phi}{\partial t^2}(a,t)+E(t)\phi(a,t)=0\, ,
\end{equation}
where $C$ and $L$ are constants, and $E(t)$ is proportional to the
(time-dependent) external magnetic flux. Equation (\ref{eq:gbc}) can be
interpreted as a sort of generalized Robin boundary condition, because of
the presence of the term with second-order time derivatives. The
theoretical analysis of that experiment was done assuming that the term proportional to the
second derivative of the field is negligible \cite{Johansson}. Whence the boundary condition
became a standard Robin boundary condition with a time dependent,
externally driven parameter.  The boundary condition that results after
implementing this approximation, has been described in terms of an
effective length~\cite{wilson1,Johansson}. 

From a theoretical point of view, the particle creation rate for the case
of a quantum field satisfying (time independent) Robin boundary conditions
on a {\it moving} mirror has been investigated in Ref.~\cite{farina2006}.
The complementary situation, namely, time dependent Robin boundary
conditions on a {\it static} mirror have been considered in
Refs.~\cite{farina2011,farina2012}.   
It is the aim of this work to consider the static and dynamical Casimir
effects for quantum fields which are subject to generalized boundary
conditions of the type defined in Eq.(\ref{eq:gbc}). Our interest in this
problem is twofold. On the one hand, the presence of the second derivative
of the field in the boundary condition modifies the usual Sturm-Liouville
problem, what manifests itself in the existence of an eigenvalue-dependent
boundary condition. This problem, well-known to mathematicians, has also
been considered in different areas of physics. It has not, however, to our
knowledge, been dealt with in the Casimir effect context. 

Regarding the experimental relevance of the inclusion of this kind of term,
we note that, although it may be neglected in the particular experimental
setup considered in Ref.~\cite{wilson1}, this contribution to the boundary
condition may indeed affect the spectral distribution of
the created particles~\cite{farina2012}, something to which future experiments 
might be sensitive.  

This paper is organized as follows. In Section~\ref{sec:classical}, we
describe the classical aspects of the model, including a mechanical
analogue, a discussion of the eigenvalue problem and the necessity of
modifying the inner product between eigenfunctions due to the presence of
the second order derivative in the boundary condition.  In
Section~\ref{sec:static} we compute the static vacuum energy,
Section~\ref{sec:dynamical} deals with the dynamical Casimir effect, and
Section~\ref{sec:conclusions} contains our conclusions.  
\section{Classical model: the loaded string analogy}\label{sec:classical}
\subsection{A mechanical analogue}
Many different boundary conditions a scalar field in $1+1$ dimensions can be
subject to, can be realized in classical mechanics analogues, based on
vibrating strings. Indeed, Dirichlet boundary conditions correspond to a
string with fixed endpoints, while Neumann conditions can be implemented by
attaching, to the corresponding endpoint,  a massless ring which can slide
freely and frictionless along a vertical rod.  Robin boundary conditions
result, in turn, when the ring is also coupled to a vertical
spring~\cite{farina2012}.  Finally, the generalized boundary condition of
Eq.(\ref{eq:gbc}) can be generated by letting the ring be massive rather
than massless (in all cases we will consider the linear regime, i.e. small
amplitude oscillations of the string). 

To see this, let us assume $T$ to be the string tension, $\mu$ its mass
density, and that its configuration may be described by a single function
$y(x,t)$, measuring its vertical departure from the equilibrium
configuration. Denoting
by $m$ the mass of the ring and by $\kappa$ the spring constant, the
position of the ring, $y(a,t)$ satisfies Newton's equation: 
 \begin{equation} 
	 m\ddot y(a,t)=-\kappa y(a,t) - T y'(a,t) \; ,
 \end{equation}
which has the same form as the generalized boundary condition of Eq.(\ref{eq:gbc}).
The vibrating string problem with this kind of condition on one of its
endpoints constitutes a well known problem in classical vibrations. 

We stress that the boundary condition for the deformation of the string is
in fact the dynamical equation for the position of the ring $q(t) \equiv
y(a,t)$, and this is the origin of the presence of second order time
derivatives in the generalized boundary condition.  A proper treatment of
the system should regard $q(t)$ and $y(x,t)$ as qualitatively different
degrees of freedom, in the sense that $q$ has a discrete, finite mass,
while $y$ is endowed with a continuous mass density. An enlightening way to
treat this kind of problem can be seen, for example,
in~\cite{Jaroszkiewicz:1989zs}, where the quantization of a non
relativistic string with an arbitrary mass distribution has been
considered. There, a system with a single continuous mass distribution
which approximates the mixed continuous and discrete case has been
considered. In this way, all the steps of the Lagrangian and Hamiltonian
formalisms, and even canonical quantization, are well defined. The desired
mass distribution is approached at the end of the process, as a special
limit, after all the stumbling blocks in the procedure are avoided.

We will follow here an alternative procedure, which, as we have explicitly
checked, yields the same results. Since the difficulties in the problem at
hand come from the fact that the discrete mass is precisely at one of the
endpoints of the system, we avoid the coincidence of those two singularities by
temporarily splitting them. Indeed, we shall first assume that  the mass
$m$ is located at an arbitrary position $x_0$, $0<x_0<a$, and impose
Neumann boundary conditions at \mbox{$x=0$} and \mbox{$x=a$}. The
generalized Robin boundary condition at $x=a$ is then recovered by taking
the `coincidence limit' $x_0\to a^-$.

The classical Lagrangian then reads:
 \begin{eqnarray}
 L&=&\frac{1}{2}\int_0^a dx\,\left[ \mu\left(\frac{\partial y}{\partial t}\right)^2-T\left(\frac{\partial y}{\partial x}\right)^2\right.\nonumber\\
 && \left. -\kappa y^2\delta(x-x_0)
 +m\left(\frac{\partial y}{\partial t}\right)^2\delta(x-x_0)\right]\;, 
 \end{eqnarray}
 with Neumann conditions implicitly assumed at \mbox{$x=0$} and \mbox{$x=a$}.

From the classical equation of motion one can easily check that the
presence of a localized mass on the string induces a discontinuity in the
spatial derivative of $y$,
\begin{equation}
T\left(\frac{\partial y}{\partial x}\vert_{x_0^+} -\frac{\partial y}{\partial x}\vert_{x_0^-}\right)=m\ddot y(x_0,t)+\kappa y(x_0,t)\, .
\end{equation}  
  
 Note that the string can interchange energy with the mass, and the
 conserved total energy of this system reads
 \begin{eqnarray}
  E&=&\frac{1}{2}\int_0^a dx\,\left[ \mu\left(\frac{\partial y}{\partial t}\right)^2+T\left(\frac{\partial y}{\partial x}\right)^2\right]\nonumber\\
 &&  +\frac{1}{2}\kappa y^2(x_0,t) +\frac{1}{2} m\left(\frac{\partial y}{\partial t}\right)_{x=x_0}^2\; ,
 \end{eqnarray}
 that is, the sum of the mechanical  energies  associated to the string and the ring.
  
\subsection{A one dimensional cavity with localized conductivity and permittivity}
Let us now focus on the analogous case of a scalar field in $1+1$
dimensions, as a toy model for the electromagnetic field in $3+1$
dimensions. We assume the Lagrangian to be given by the expression:
 \begin{eqnarray}
 L&=&\frac{1}{2}\int_0^a dx\,\left[ \epsilon(x,t)\left(\frac{\partial \phi}{\partial t}\right)^2-\left(\frac{\partial \phi}{\partial x}\right)^2\right.\nonumber\\
&&\left. - V(x,t) \phi^2\right ]\, ,
 \label{model2}
 \end{eqnarray}
 with 
 \begin{equation}
\epsilon(x,t)=1+\chi(t)\delta(x-x_0)\, ,
 \end{equation}
 and 
  \begin{equation}
 V(x,t)=v(t)\delta(x-x_0)\, .
 \end{equation}
 As in the mechanical model, we shall regard this as a simple model to
 describe a cavity in which the permittivity and conductivity are
 concentrated at the  point $x=x_0$, which, tending to $a$ from the left,
 reproduces the generalized Robin condition at $x=a$. The particular case
 $\chi(t)=0$ and $v(t)=v_0(1+f(t))$ has been considered in
 Ref.\cite{crocce}, as a simple model of the experimental setup of
 Ref.\cite{padova} (the generalization to the electromagnetic case has been
 analyzed in Ref.\cite{japon}). In Section \ref{sec:static} we will compute
 the Casimir interaction energy between two slabs described by constant values 
of $v=v_0$ and $\chi=\chi_0$. In Section IV we will compute the photon creation associated to
a time dependent $v(t)$ and $\chi=\chi_0$.

\subsection{Eigenfunctions and inner product}
 Let us consider the model given in Eq.(\ref{model2}), with Neumann
 boundary conditions at $x=0,a$. We will compute the eigenfunctions and
 eigenvalues for the particular case $v(t)=v_0$ and $\epsilon(t)=\chi_0$.  The eigenmodes can be
 written as
 \begin{eqnarray}
&& \Psi_k(x,t)=N_ke^{-ikt}\left[ \cos(kx)\cos(k(x_0-a))\theta(x_0-x)\right.\nonumber\\
 &&\left.+ \cos(kx_0)\cos(k(x-a))\theta(x-x_0)\right]\equiv e^{-ikt} \psi_k(x)\, ,
 \end{eqnarray} 
 where $N_k$ is a normalization constant. 
They are continuous at $x=x_0$ and satisfy Neumann boundary conditions at
$x=0,a$. The discontinuity equation at $x=x_0$ implies
 \begin{equation}
 k\sin(ka)=(v_0-\chi_0 k^2) \cos(kx_0)\cos(k(x_0-a))\, ,
 \label{gbcx0}
 \end{equation}
 which is the equation that defines the eigenfrequencies. In the particular
 case $x_0\to a$,  the transcendental equation that defines the
 eigenfrequencies  simplifies to
  \begin{equation}
 k\tan(ka)=(v_0-\chi_0 k^2) \,  .
 \label{gbcmodel}
 \end{equation}  

 It is straightforward to show that, unless $\chi_0=0$, eigenfunctions
 corresponding to different eigenvalues are not orthogonal with the usual
 inner product. However, defining a generalized inner product (see the
 Appendix and Refs.~\cite{Jaroszkiewicz:1989zs,darma})
 \begin{eqnarray}
 (\psi_k,\psi_{k'})&=&\int_0^a dx \, \psi_k(x)\psi_{k'}(x)+\chi_0\psi_k(a)\psi_{k'}(a) \nonumber\\
 &=& \int_0^a dx \, \epsilon(x)\psi_k(x)\psi_{k'}(x) \label{inner}
 \end{eqnarray}
 one can check the orthogonality $(\psi_k,\psi_{k'})=0$ for $k\neq k'$. With appropriate normalization,
 the eigenfunctions can be chosen to be orthonormal, as we shall assume it
 has been done in what follows. 

 This phenomenon is a general feature of the hybrid continuous
 plus discrete systems, where one can show that the eigenfunctions of the
 Hamiltonian are orthogonal for a scalar product defined in terms of a
 kernel~\cite{Jaroszkiewicz:1989zs}, which defines a Sturm-Liouville
 problem.
 
 The equations above are valid  even in the time-dependent case $v_0\to v(t)$, in which the 
 eigenvalues become parametrically dependent on time, as well as the eigenfunctions $\psi_k$.  We will analyze the time-dependent situation in Section  IV.
 
 Writing the field $\phi$ as a linear combination of the spatial eigenfunctions
 \begin{equation}
 \phi(x,t)=\sum_k Q_k(t)\psi_k(x)\, ,
 \end{equation}
  and inserting this expression into the classical Lagrangian one can check
  that it reduces to a set of uncoupled harmonic oscillators $Q_k$, with frequency $k$:
 \begin{equation}
  L=\frac{1}{2}\sum_k(\dot Q_k^2-k^2Q_k^2) \;.
  \end{equation} 
Details of this calculation are presented in the Appendix. Note that the
additional term in the inner product Eq.(\ref{inner}) is crucial to  cancel
the kinetic term concentrated at $x_0$.  Note also that these results can
be straightforwardly generalized to cases where several slabs are located 
at different positions.
\section{Static Casimir effect}\label{sec:static}
Once the system has been reduced to a set of uncoupled harmonic oscillators, the
calculation of the vacuum energy can be performed by direct mode-summation. 
We will explicitly compute the static Casimir effect for two different physical 
situations: the interaction between two thin slabs, each one  described by
its conductivity and permittivity (in free space), 
and a system that satisfies generalized boundary conditions.  

\subsection{Interaction vacuum energy for two slabs }
Let us consider two slabs, located at $x=\pm a/2$, and described by
permittivities $\chi_0^\pm$ and conductivites $v^\pm$, respectively. In
order to have a discrete set of eigenfrequencies, we enclose the system in
a box of size $2L$, and impose Neumann boundary conditions at $x=\pm L$. As
we will take the limit $L\gg a$ at the end of the calculation, the result
will be independent of $L$ and of the boundary condition imposed at $x=\pm
L$.

The eigenfunctions that satisfy Neumann boundary conditions at $x=\pm L$ can be written as 
\begin{equation}
f_\omega(x) = \left\{
\begin{array}{cl}
	A_1 \cos[\omega(x+L)] & {\rm for} \, -L<x<-\frac{a}{2} \\
	A_2 \cos(\omega x) + A_3 \sin(\omega x) & {\rm for} \,
	-\frac{a}{2}<x<\frac{a}{2}\\
	A_4 \cos[\omega(x-L)]  & {\rm for} \, \frac{a}{2}<x<L \, .
\end{array}
\right.
\label{fw}
\end{equation}
The function $f_\omega(x)$ must be continuous at $x=\pm a/2$ and the spatial
derivatives must satisfy:
\begin{equation}
{\rm disc} [\partial_x f_\omega]_{x=\pm a/2}=[v_0^{\pm}-\chi_0^{\pm}\omega^2]f_\omega(\pm \frac{a}{2})\, .
\label{disc}
\end{equation}
The eigenfrequencies are the solutions of $\det M=0$, where $M(a,L,\omega)$  is the $4\times 4$ matrix associated to the linear system
of equations for the coefficients $A_i$ derived by inserting Eq.(\ref{fw}) in Eq.(\ref{disc}).  After some straightforward calculations one can show that
\begin{eqnarray}
&&\det[M(a,L,\omega)]=\Delta^+  \Delta^- (\sin (2 \omega (a-L))+2 \sin (a \omega))+\nonumber\\
&& 2 (\Delta^+ +\Delta^- ) \cos (a \omega)+(\Delta^+  \Delta^--4) \sin (2 L \omega)+\nonumber\\
&& 2 (\Delta^+ +\Delta^-)  \cos (2 L \omega)\,\, ,
   \end{eqnarray}
where $\Delta^\pm=\frac{1}{\omega}(v_0^\pm-\chi_0^\pm\omega^2)$.

We will compute the Casimir energy as the difference between the zero point
energy of the slabs separated by a distance $a$, and  that corresponding to
a distance $l\gg a$. Using the argument theorem, we see that:
\begin{eqnarray}
  E_C(a)&=&\frac{1}{2}\sum_n(\omega_n-\tilde\omega_n)\nonumber \\
  &=&-\frac{1}{4\pi i}\oint
  dz\, \log \det\left[\frac{M(a,L,z)}{M(l,L,z)}\right]\, ,
  \label{form0}
\end{eqnarray}
where $\omega_n$ and $\tilde\omega_n $ are the eigenfrequencies associated
to the distances $a$ and $l$, respectively.  The integration path must
include the real positive axis. Following standard steps, and taking the
limit $L\to\infty$, we arrive to an integral in the imaginary-frequency
axis $z=i\xi$:
\begin{equation}
	E_C=\frac{1}{2\pi }\int_0^\infty d\xi\, \log\big[1-e^{-2\xi
	a}C(\xi)\big]\; ,
\end{equation}
where
\begin{equation}
C(\xi)= \frac{(v_0^++\chi_0^+\xi^2)(v_0^-+\chi_0^-\xi^2)}{(2\xi+v_0^++\chi_0^+\xi^2)(2\xi+v_0^-+\chi_0^-\xi^2)}\, .
\end{equation}
Not surprisingly, for the particular case $\chi_0^{\pm}=0$, this result
coincides with the usual Casimir energy for the so called $\delta$-potentials.
Moreover, in the $v_0^\pm\to \infty$ limit, one has  $C(\xi)\to 1$, and the
result reproduces the usual one for Dirichlet boundary conditions.
Note that, as $\partial C/\partial \chi_0^\pm >0$,  the  presence of the second order time derivative
in the boundary conditions enhances the interaction between slabs. 


It is interesting to remark that the final result for the Casimir energy is
tantamount to the one corresponding to a $\delta$-potential with a
frequency-dependent coefficient $v(\omega)=v_0-\chi_0\omega^2$.
Therefore, this result could have been derived using Lifshitz formula with
the particular reflection coefficients that describe the slabs. The above
calculation is an alternative and equivalent way to compute the vacuum energy, that shows that
the Casimir energy is just the sum over the eigenfrequencies defined by the
$\omega$-dependent boundary conditions.  

\subsection{Interaction vacuum energy for generalized boundary conditions}
We now consider the second case, namely, a system that satisfies Neumann
boundary conditions at $x=0$ and generalized Robin boundary conditions at
$x=a$. The eigenfrequencies are defined implicitly by  Eq.(\ref{gbcmodel}),
that we rewrite as $G(a,\omega)=0$, with
\begin{equation}
G(a,\omega)=\omega\sin(\omega a)-(v_0-\chi_0\omega^2)\cos(\omega a)\, .
\end{equation}
As in the previous subsection, we enclose the system in a large box of size
$2L$, and impose Neumann boundary conditions at $x=\pm L$. The
eigenfrequencies are thus determined by the equation $\tilde
G(a,L,\omega)=0$, with

\begin{figure}[h!]
\includegraphics[width=\linewidth]{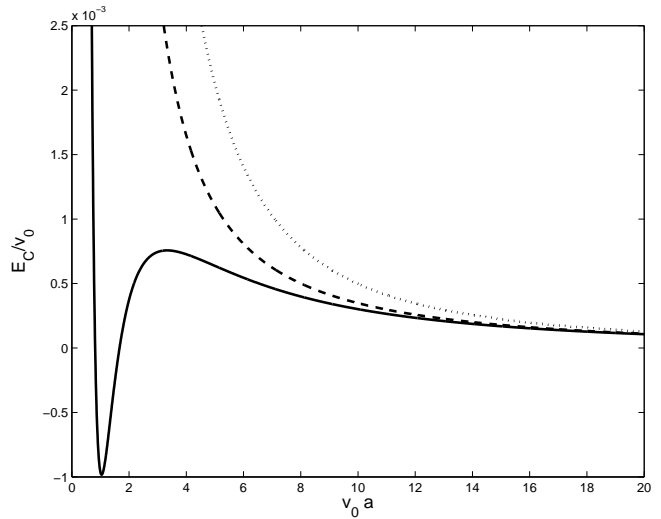}
\caption{Dimensionless total Casimir energy $E_C/v_0$  as a function of the distance
	$v_0 a$, for  different values of
	$v_0\chi_0$. While the force is always repulsive
	for large values of $v_0\chi_0$, for $v_0\chi_0=0.1$ (solid line) it
	starts being repulsive, then becomes attractive, and
	is finally repulsive again for larger values of the distance. Dashed line corresponds to $v_0\chi_0= 0.5$, 
and the dotted line corresponds to $v_0\chi_0=5$}
\end{figure}

\begin{equation}
\tilde G(a,L,\omega)= G(a,\omega)\sin[\omega(L-a/2)]^2\, .
\end{equation}
Then we compute the Casimir energy as the difference between the vacuum energy associated to the
length $a$, and that associated to $l$ with $l\gg a$. Using again the argument theorem
  \begin{equation}
  E_C(a)=-\frac{1}{4\pi i}\oint dz\, \log \left[\frac{\tilde G(a,L,z)}{\tilde G(l,L,z)}\right]\, .
  \label{form}
  \end{equation}
the final result can be written as 
\begin{equation}
  E_C=\frac{1}{2\pi }\int_0^\infty d\xi\, \log(1-e^{-2\xi a}B(\xi))\, ,
  \label{casimirgrbc}
  \end{equation}
  where the function $B$ contains the information about the generalized boundary condition 
  \begin{equation}
  B(\xi)=\frac{\xi-v_0-\chi_0\xi^2}{\xi+v_0+\chi_0\xi^2}\, .
  \end{equation}  
As the sign of $B$ depend on the values of $v_0$, $\chi_0$ and $\xi$, the force can be attractive or repulsive
depending on the value of $a$.  

In oder to analyze the behavior of the energy with the different parameters, it is
useful to note that
\begin{equation}
\frac{E_C}{v_0}=h(v_0 a,v_0\chi_0)\, ,
\label{adim}
\end{equation} where $h$ is a dimensionless function. This can be easily checked changing
variables
 $\xi\to\xi/v_0$ in Eq.(\ref{casimirgrbc}).
Therefore, the qualitative dependence of the energy with the distance  only depends on the 
dimensionless quantity $v_0\chi_0$. For instance, when $v_0\chi_0>1/4$, $B$ is negative for all values of
$\xi$, and therefore the force is repulsive.
In particular, in the limit $v_0\to\infty$, the boundary condition at
$x=a$ becomes Dirichlet boundary condition, and  one has
$B\to -1$.  The Casimir energy becomes the standard result for a scalar field satisfying
Neumann boundary conditions at $x=0$ and Dirichlet boundary conditions at
$x=a$, which corresponds to a repulsive force. When $v_0\chi_0<1/4$,
we cannot predict the sign of the force analytically.  

In Fig. 1  we present some numerical evaluations of
Eq.(\ref{casimirgrbc}). We refer all the ingredients in the expression for the energy to the 
dimensionful quantity, $v_0$.
We plot  the Casimir dimensionless energy $E_C/v_0$ as a function of the distance
$a$ (in units of $v_0$) for different values of $v_0\chi_0$.
We see that for small distances, the force changes sign  for 
the smaller value of  $v_0\chi_0=0.1$. 
In this regime, the force cross from repulsive to attractive  
and back sign to repulsive again, as the distance increases. For other values of 
$v_0\chi_0>1/4$ the force is, as expected, always repulsive.

\section{Dynamical Casimir effect}\label{sec:dynamical}
Let us now consider the case of a time dependent $v(t)=v_0(1 + f(t))$ and
constant $\chi(t)=\chi_0$.  In order to analyze this problem one can
proceed as usual \cite{reviews dce,crocce,trembling}, showing that, at the classical
level, the system can be described by a set of coupled harmonic oscillators
with time dependent frequencies and couplings.
To this end, we introduce an `instantaneous basis' $\psi_k(x,t)$ through the equations
\begin{eqnarray}
\psi_k'' +k^2(t)\psi_k&=&0\nonumber\\
\psi_k'(0)&=&0\nonumber\\
\psi_k'(a)&=&0\nonumber\\
\psi_k'(x_0^+)-\psi_k'(x_0^-)&=&(-\chi_0 k^2+v(t))\psi_k\, .
\end{eqnarray}
Note that, due to the time dependent boundary condition at $x_0$,  the eigenvalues are time dependent, and 
this induces a parametric time dependence in the basis functions. The limit
$x_0\to a^-$ describes a cavity with a SQUID at one end, but one could
consider more general situations like a cavity ended by two SQUIDs, or even
a set of SQUIDs located at different positions in a waveguide.

We now expand the field using the instantaneous basis
\begin{equation}
 \phi(x,t)=\sum_k Q_k(t)\psi_k(x,t)\, .
 \label{expinst}
 \end{equation}
The classical Lagrangian given in Eq.(\ref{model2}) can be written in terms
of the variables $Q_k$, as in the static case. The frequencies of the
classical oscillators become time-dependent, and the Lagrangian contains
additional terms proportional to derivatives of the basis functions, which
in turn are proportional to the derivatives of the eigenvalues $\dot k(t)$.

As shown in the Appendix, the classical Lagrangian reads
\begin{eqnarray}
 L&=&\frac{1}{2}\sum_k(\dot Q_k^2-k^2(t)Q_k^2)+\sum_{kj}A_{kj}(t)\dot Q_kQ_j\nonumber\\
 &-&\frac{1}{2}\sum_{kj}S_{kj}(t)Q_kQ_j\label{Ldet}\, ,
 \end{eqnarray}
where the time-dependent matrices $A_{kj}$ and $S_{kj}$ can be chosen to be
antisymmetric and symmetric, respectively. The explicit expressions for
them are derived in the Appendix. Note that, if the time dependence of the eigenvalues
is proportional to a dimensionless parameter $\delta$, $A_{kj}=O(\delta)$ and  $S_{kj}=O(\delta^2)$
(see Eq.(\ref{ays})).

Let us now assume that the externally driven property  $v(t)$ 
has a harmonic time dependence with frequency $\Omega$. When the external
frequency is tuned with an eigenfrequency of the static cavity
$\Omega=2\tilde k_0$, with $\tilde k_0$ one of the solutions to Eq.(\ref{gbcx0}), we
expect the number of created photons to be enhanced by parametric
resonance.  Moreover, since in general the spectrum of the cavity is not
regularly-spaced, it is reasonable to neglect couplings between modes~\cite{detuning}. 
The dynamics of the system is essentially given by that of the mode
associated to $k_0$,  i.e. a single harmonic oscillator with time dependent
frequency:
\begin{equation}
 L\simeq\frac{1}{2}[\dot Q_{k_0}^2-k_0^2(t)Q_{k_0}^2]\label{MSA}\, ,
\end{equation}
where 
we have neglected terms of order $O(\delta^2)$. Here 
$k_0(t)$ is a solution to 
 \begin{equation}
 k\sin(ka)=(v(t)-\chi_0 k^2) \cos(kx_0)\cos(k(x_0-a))\, .
 \label{gbcx0det}
 \end{equation}
 
 For the sake of simplicity, we will solve this equation for $x_0\to a^-$.
 We denote by $\tilde k_0$ a solution to Eq.(\ref{gbcx0det}) in the static case $f(t)=0$. Assuming that $k_0(t)=\tilde k_0(1+\eta f(t))$, with $\eta f(t)\ll 1$, 
 one can show that
 \begin{equation}
 \eta=\frac{v_0}{v_0+\tilde k_0^2(a+\chi_0)+ a (v_0-\chi_0\tilde k_0^2)^2}\, .
 \end{equation}
 
When $f(t)=A \sin[2\tilde k_0 t]$, the number of photons with frequency $\tilde k_0$ will grow
exponentially. The calculation is, by now, standard~\cite{reviews dce,crocce,trembling,detuning},
and will not be reproduced here.  The result is
\begin{equation}
N_{\tilde k_0}(t)=\exp[ {\tilde k}_0 A \eta t]\equiv \exp[\lambda t]\, .
\end{equation}
Note above equations are valid for small values of the aforementioned parameter $\delta=A\eta\ll 1$.
  
It is interesting to analyze two opposite limiting cases. When $v_0 a\gg 1$, 
the lowest frequency solution is $\tilde k_0\simeq \pi/(2 a)$, and therefore $\eta\simeq1/(v_0 a)$. The particle creation 
rate, in this case, is $\lambda = \pi A/(2 v_0 a^2)$. 

On the other hand,  when $v_0 a \ll 1$,   the first solution to the transcendental Eq. (\ref{gbcmodel}) is 
\begin{equation}
\tilde k_0 a \simeq\sqrt{\frac{v_0 a}{1+\frac{\chi_0}{a}}}\, .
\end{equation} In this case we have $\eta \approx 1/2$ (assuming that $\chi_0/a\ll 1$). Therefore, the 
corresponding rate is $\lambda= A/2 \sqrt{v_0/a}$.

The last case may be of some interest for the experimental observation of the dynamical Casimir effect, since the Robin or generalized Robin
boundary conditions may be adjusted in such a way to reduce the value of the lowest eigenfrequency of the unperturbed cavity.
This point deserves further analysis. 

When considering the parametric resonance situation, it is crucial to tune
the external frequency with one of the eigenfrequencies of the system. The
term proportional to $\chi_0$ in the boundary condition, however small,
does introduce significant modifications to the static eigenfrequencies of the 
cavity. This effect may be relevant, for instance,
for an experiment with a coplanar waveguide of finite size.
  
As an illustration of the last point, let us assume that $x_0\to a^-$ and that the external
frequency is twice the lowest eigenfrequency of the cavity (that is, twice the first  
solution to Eq.(\ref{gbcmodel})). In a realistic situation~\cite{wustmann}, $v_0
a\gg 1$, and an approximate solution is $\tilde k_0\simeq\pi/(2a)$.
Expanding the transcendental equation around this solution we obtain
 \begin{equation}
 {\tilde k}_0\simeq \frac{\pi}{2a}(1-\frac{1}{v_0 a} + \frac{1}{v_0^2 a^2} - \frac{\pi^2}{4} \frac{\chi_0}{v_0^2 a^3})\, .
 \end{equation}
 Assuming \cite{wustmann} $v_0 a =20$ and $\chi_0/a=0.05$ the
 correction to the lowest eigenfrequency due to $\chi_0$ is $3\times
 10^{-4}$.  This small correction should be taken into account in order to
 achieve parametric resonance if the amplitude of the time-dependent
 external conditions is sufficiently small~\cite{detuning}.

\section{Conclusions}\label{sec:conclusions}
The original aim of this work has been to analyze the static and dynamical Casimir
effects when the field satisfies generalized boundary conditions involving
second order time derivatives.  These conditions were in turn motivated by
the effective boundary condition satisfied by the magnetic flux in a
waveguide terminated by a SQUID, a setup that has been recently employed as
a device to measure the creation of photons from the vacuum in the presence of
time-dependent external fields~\cite{wilson1}. 

We have shown that this problem does have a simple and well-known classical
analogue: a loaded string. From this point of view, the presence of the
second-order time derivative is not surprising, since the boundary
condition is in this case nothing but the dynamical equation for a massive
ring attached to the end of the string. From a mathematical point of view,
we have eigenvalue-dependent boundary conditions, and therefore the
eigenfunctions associated to different eigenvalues are not orthogonal under 
the usual inner product. A generalization of the inner product makes them
orthogonal.  When expanding the deformation of the string in terms of
spatial eigenfunctions, the additional term in the inner product exactly
cancels out the kinetic term associated to the ring, and one ends up with a set of
decoupled harmonic oscillators. 

This mechanical  analogy lead us to consider a more general situation, in
which the ring is not attached to the end of the string; rather, it is
located at an arbitrary distance from the endpoint.  From a field-theory
perspective, this can be considered as a toy model for an electromagnetic cavity in which one
inserts a thin slab characterized by its conductivity and permittivity.
From the quantum circuits point of view, this corresponds to a situation in
which a SQUID is inserted in a one-dimensional waveguide.

Therefore, we computed the static Casimir interaction energy between two
slabs, generalizing previous results for $\delta$-potentials.  We also
computed the static vacuum energy for the particular case in which the slab
is near a border of the cavity. In summary,  we have shown that the Casimir energy can be computed as
the sum over modes satisfying the generalized Robin boundary
condition.

Finally, and coming back to the original motivation, we considered some
particular aspects of the dynamical Casimir effect for the generalized
boundary conditions. On the one hand, we computed the particle creation rate
assuming parametric resonance for the case of a finite waveguide
ended by a SQUID. On the other hand,
we discussed
 the influence of the second order time derivative on
the tuning of the external pumping with one
of the eigenfrequencies of the cavity. We have seen that the term proportional to $\ddot\phi$
in the boundary condition may indeed be relevant for this tuning. 

The analysis of loaded strings with masses distributed periodically along
it, or analogue acoustic and elastic systems,  induces the presence of
(approximate) band gaps in the spectrum~\cite{bands}.  It would be
interesting to analyze theoretically electromagnetic analogues of these
configurations, like a waveguide with several  SQUIDs, distributed
periodically on it, or a microwave cavity with several slabs inserted
accordingly.

\appendix
\section{} \label{A}

In this Appendix we present some details of the calculations of the
classical Lagrangian in both the static and dynamical cases. 

To begin with, let us show that it is necessary to modify the usual inner
product in order to have an orthonormal basis. 
The eigenfunctions satisfy
\begin{equation}
\psi_k'' +k^2\psi_k=0
\end{equation}
with the boundary conditions
\begin{eqnarray}
\psi_k'(0)&=&0\nonumber\\
\psi_k'(a)&=&0\nonumber\\
\psi_k'(x_0^+)-\psi_k'(x_0^-)&=&(-\chi_0k^2+v_0)\psi_k\, .
\end{eqnarray}
For the sake of definiteness, we choose Neumann boundary conditions at $x=0,a$. The results below can be generalized to the case of Robin boundary conditions.

We compute
 \begin{equation}
 I_{ij}=\int_0^a(\psi_i''\psi_j-\psi_i\psi_j'') dx=(j^2-i^2)\int_0^a\psi_i\psi_jdx \, , 
 \label{inner1}
 \end{equation}
 where in the last equality we used the eigenvalue equation.
 On the other hand we have
  \begin{equation}
 I_{ij}=\int_0^a[(\psi_i'\psi_j)'-(\psi_i\psi_j')') dx=\chi_0(i^2-j^2)\psi_i(x_0)\psi_j(x_0)\, .
 \label{inner2}
 \end{equation}
 One should be careful with the evaluation of 
 \begin{equation}
 \int_0^a (\psi_i\psi_j')' dx
 \end{equation}
 because of the discontinuity at $x_0$. Eq.(\ref{inner2}) can be obtained writing $\int_0^a=\int_0^{x_0^-}+\int_{x_0^+}^a$ 
and using the boundary condition at $x_0$:
\begin{eqnarray}
 \int_0^a (\psi_k\psi_j')' dx &=&\psi_k(x_0)[\psi_j'(x_0^-)-\psi_j'(x_0^+)]\nonumber\\
 &=& (\chi_0j^2-v_0)\psi_k(x_0)\psi_j(x_0)\label{warn} 
 \end{eqnarray}  
 Subtracting Eqs.(\ref{inner1}) and (\ref{inner2}) we get
\begin{equation}
0=(j^2-i^2)\left [\chi_0\psi_i(x_0)\psi_j(x_0)+\int_0^a dx \psi_i\psi_j\right] \, ,
\end{equation}
 and therefore the inner product defined as 
 \begin{eqnarray}
 (\phi_i,\phi_j)&=&\chi_0\psi_i(x_0)\psi_j(x_0)+\int_0^a dx \, \psi_i\psi_j\nonumber\\
&=& \int_0^a dx \, \epsilon(x)\psi_i\psi_j\ \end{eqnarray}
 vanishes when $i\neq j$.
 
 To compute the static classical Lagrangian we write
 \begin{equation}
 \phi(x,t)=\sum_k Q_k(t)\psi_k(x)\, .
 \end{equation}
 Therefore
  \begin{eqnarray}
 \int_0^a d x\, \dot\phi^2&=&\sum_{k j} \dot Q_k\dot Q_j\int_0^a dx\,\psi_k\psi_j\nonumber\\
&=& \sum_{k j} \dot Q_k\dot Q_j(\delta_{kj}-\chi_0\psi_k(x_0)\psi_j(x_0))\nonumber\\
&=& \sum_k\dot Q_k^2-\chi_0\dot\phi^2(x_0)
\label{tempderiv}
\end{eqnarray} 

A similar calculation can be done for the spatial derivatives
 \begin{eqnarray}
 \int_0^a d x\phi'^2&=&\sum_{k j}  Q_kQ_j\int_0^a dx\psi_k'\psi_j'\nonumber\\
&=& \sum_{k j} Q_kQ_j\int_0^a[ (\psi_k\psi_j')'-\psi_k\psi_j'')]\nonumber\\
&=& \sum_{k j} Q_kQ_j\int_0^a[ (\psi_k\psi_j')'+j^2\psi_k\psi_j)]
\label{spatialderiv}
\end{eqnarray} 
 Inserting Eq.(\ref{warn}) in (\ref{spatialderiv}), and using again the orthogonality
 \begin{equation}
  \int_0^a d x\phi'^2=\sum_kk^2Q_k^2-v_0\phi^2(x_0,t)\, .
  \label{spatialderiv2}
  \end{equation}
  Using Eqs.(\ref{tempderiv}) and (\ref{spatialderiv2}), the classical Lagrangian Eq.(\ref{model2}), can be written
  in the static case as
  \begin{equation}
  L=\frac{1}{2}\sum_k(\dot Q_k^2-k^2Q_k^2)\, .
  \end{equation} 
 
 We now consider the time dependent situation $v_0\to v(t)$.  Using the basis functions introduced in Section IV, 
from Eq.(\ref{expinst}) we have
\begin{equation}
\dot \phi(x,t)=\sum_k (Q_k(t)\dot \psi_k(x,t)+\dot Q_k(t) \psi_k(x,t)) \, ,
\end{equation}
 and therefore
 \begin{eqnarray}
\frac{1}{2}\int_0^a dx\, \epsilon(x)\dot\phi^2 &=& \frac{1}{2}\sum_k \dot Q_k^2+\sum_{kj}\dot Q_k Q_j A_{kj}\nonumber\\
&-&\frac{1}{2} \sum_{kj} Q_k Q_j S_{kj}\, , 
\end{eqnarray}
 where
 \begin{eqnarray}
 A_{kj}&=&\int_0^a dx\, \epsilon(x)\, \psi_k\dot\psi_j\nonumber\\
 S_{kj}&=&-\int_0^a dx\, \epsilon(x)\, \dot\psi_k\dot\psi_j\, . 
 \label{ays}
 \end{eqnarray}
 Note that, due to the orthogonality of the eigenfunctions,  the matrix $A_{kj}$ is antisymmetric. 
 The matrix $S_{kj}$ is obviously symmetric.

\section*{Acknowledgements}
We would like to thank C. Farina for useful discussions. This work was supported by ANPCyT, CONICET, UBA and UNCuyo.


\begin{thebibliography}{bib}
\bibitem{libros}
P.W. Milonni, The Quantum Vacuum (Academic Press, San Diego, 1994); M.
Bordag, U. Mohideen, and V.M. Mostepanenko, Phys. Rep. {\bf 353}, 1 (2001);
K. A. Milton, The Casimir Effect: Physical Manifestations of the Zero-
Point Energy (World Scientific, Singapore, 2001); S. Reynaud, A. Lambrecht,
C. Genet, and M.T. Jaekel, et al., C. R. Acad. Sci. Paris Ser. {\bf IV 2},
1287 (2001); K. A.	Milton,	J.	Phys.	A	{\bf 37},
R209	(2004);	S. K. Lamoreaux, Rep. Prog. Phys. {\bf 68}, 201 (2005); M.
Bordag,	G. L.	Klimchitskaya,	U.	Mohideen,	and	V. M.
Mostepanenko, Advances in the Casimir Effect (Oxford University Press,
Oxford, 2009).
\bibitem{reviews dce}For recent reviews see
V.~V.~Dodonov,
  Phys.\ Scripta {\bf 82} (2010) 038105; 
 D.~A.~R.~Dalvit, P.~A.~Maia Neto and F.~D.~Mazzitelli,
  Lect.\ Notes Phys.\  {\bf 834} (2011) 419. 
\bibitem{wilson1}C.M. Wilson et al, Nature {\bf 479}, 376 (2011).
\bibitem{padova}A. Agnesi,  {\it  et al}, J. Phys. A: Math. Gen. {\bf 41}, 164024 (2008).
\bibitem{Kim06}  W.~-J.~Kim, J.~H.~Brownell and R.~Onofrio,
  Phys.\ Rev.\ Lett.\  {\bf 96}, 200402 (2006).
  \bibitem{Johansson}J. R. Johansson, G. Johansson, C. M. Wilson, and Franco Nori
Phys. Rev. A{\bf  82}, 052509 (2010).
\bibitem{farina2006}B. Mintz, C. Farina, P. A. Maia Neto and R. B. Rodrigues, J. Phys. A: Math. Gen. {\bf 39}, 6559 (2006);
A.~L.~C.~Rego, B.~W.~Mintz, C.~Farina and D.~T.~Alves,
  Phys.\  Rev.\  D 87, {\bf 045024} (2013).
\bibitem{farina2011} H.O. Silva and C. Farina, Phys. Rev. D{\bf 84}, 045003 (2011).
\bibitem{farina2012} C. Farina et al, Int. J. Mod. Phys. Conf. Ser. {\bf 14}, 306 (2012).
\bibitem{Jaroszkiewicz:1989zs} G.~A.~Jaroszkiewicz and H.~Perry, J.\ Phys.\ A {\bf 23}, 3621 (1990).
\bibitem{crocce}
M. Crocce, D.A.R. Dalvit, F.C.  Lombardo, and F.D.  Mazzitelli, Phys. Rev. A {\bf 70}, 033811 
(2004).
\bibitem{japon}
W. Naylor,  S.  Matsuki, T. Nishimura, and  Y. Kido, Phys. Rev. A {\bf 80}, 043835 (2009).
\bibitem{darma}Darmawijoyo and W.T. Van Horssen, Journal of Vibration and Control {\bf 9}, 1231 (2003).
\bibitem{trembling}R.~Schutzhold, G.~Plunien and G.~Soff,
  Phys.\ Rev.\ A {\bf 57}, 2311 (1998).
  \bibitem{wustmann} W. Wustmann and V. Shumeiko, arXiv 1302.3484
\bibitem{detuning}  See for instance   M.~Crocce, D.~A.~R.~Dalvit and F.~D.~Mazzitelli,
  Phys.\ Rev.\ A {\bf 64}, 013808 (2001).
  \bibitem{bands}
  J.P Dowling, Journal Acoustic Soc. Am. {bf 91}, (1992);   
  M.S: Kushwaha, P.  Halevi, L. Dobrzynski and B. Djafari-Rouhani, Phys. Rev. Lett. {\bf 71}, 2022 (1993).  
  
  \end{thebibliography}
\end{document}